\documentclass[twocolumn]{webofc}
\usepackage[varg]{txfonts}   % Web of Conferences font
\usepackage{amssymb}
\usepackage{graphicx}
\usepackage{dcolumn}
\usepackage{bm}
\usepackage{epsfig}
\usepackage{amssymb}
\usepackage{amsmath}
\usepackage{color}
\usepackage{datetime}
\usepackage{wasysym}
\usepackage{slashed}
\usepackage[hidelinks]{hyperref} 
\usepackage{microtype}

\newcommand{\bal}{\begin{align}}
\newcommand{\eal}{\end{align}}
\newcommand{\beq}{\begin{eqnarray}}
\newcommand{\eeq}{\end{eqnarray}}
\newcommand{\nneeq}{\nonumber \end{eqnarray}}

\newcommand{\cB}{ {\cal B} }

\newcommand{\cM}{ {\cal M} }
\newcommand{\cH}{ {\cal H} }
\newcommand{\cE}{ {\cal E} }

\newcommand{\cG}{ {\cal G} }
\newcommand{\cK}{ {\cal K} }

\newcommand{\cV}{ {\cal V} }

\newcommand{\cX}{ {\cal X} }

\newcommand{\cU}{ {\cal U} }

\newcommand{\cY}{ {\cal Y} }

\newcommand{\bmat}{\left[\begin{array}}
\newcommand{\emat}{\end{array}\right]}

\usepackage[utf8]{inputenc}

\begin{document}
\title{Asymptotic freedom using a gluon mass as a regulator}

\author{\firstname{Juan José} \lastname{Gálvez-Viruet}\inst{1}\fnsep\thanks{\email{jj.galvezviruet@ugr.es}}
    \and
    \firstname{María} \lastname{Gómez-Rocha}\inst{1,2}\fnsep\thanks{\email{mgomezrocha@ugr.es}} 
}

\institute{Departamento de Física Atómica, Molecular y Nuclear, Universidad de Granada, E-18071 Granada, Spain.
\and
%Departamento de Física Atómica, Molecular y Nuclear and 
Instituto Carlos I de Física Teórica y Computacional, Universidad de Granada, E-18071 Granada, Spain.}

\abstract{%
Front-Form Hamiltonian dynamics provides a framework in which QCD's vacuum is simple and states are boost invariant. However, canonical expressions are divergent and must be regulated in order to establish well-defined eigenvalue problems. The Renormalization Group Procedure for Effective Particles (RGPEP) provides a systematic way of finding counterterms and obtaining regulated Hamiltonians. Among its achievements is the description of asymptotic freedom, with a running coupling constant defined as the coefficient in front of the three gluon-vertex operators in the regulated Hamiltonian. However, the obtained results need a deeper understanding, %is not utterly understood, 
since the coupling exhibits a finite dependence on the regularization functions, at least at the third-order term in the perturbative expansion. Here we present a similar derivation using a different regularization scheme based on massive gluons. The procedure can be extended to incorporate contributions from virtual fermions.
}
\maketitle
%%%%%%%%%%%%%%%%%%%%%%
\section{Introduction}
\label{Intro}
%%%%%%%%%%%%%%%%%%%%%%
Front-Form Hamiltonian dynamics~\cite{dirac_forms_1949,brodsky_quantum_1998} is a candidate tool to characterize bound states in QCD  \cite{gomez-rocha_asymptotic_2017,glazek_renormalized_2017} and to investigate the relation between the parton and constituent quark models aiming to obtaining results that are invariant under certain boost transformations \cite{wilson_nonperturbative_1994}. However, these long-term goals have important challenges to overcome. One of them is the regularization of highly divergent canonical expressions. Another one is the introduction of counterterms to describe aspects related to vacuum physics~\cite{wilson_nonperturbative_1994}. In this context, the similarity renormalization group, developed by Głazek and Wilson~\cite{glazek_renormalization_1993,glazek_perturbative_1994}, together with the concept of effective particle introduced by Głazek~\cite{glazek_similarity_1997,glazek_perturbative_2012,glazek_effective_2017}, known as RGPEP, stands for a systematic procedure to handle these divergences and to find counterterms.

The RGPEP is in a developing stage and the way to obtain non-perturbative solutions to the renormalization-group equation is still unknown. However, it is possible to use perturbative expansions in powers of the coupling constant instead~\cite{glazek_perturbative_2012}. The bound state equation has been considered in heavy-flavor QCD and numerical results for the spectrum of heavy quarkonia and baryons have been obtained using a simplified sketch~\cite{Glazek:2017rwe,Serafin:2018aih}. Initially, the new version of the method was used to describe the running coupling and, more precisely, the phenomenon of asymptotic freedom. 
Published works in this direction~\cite{glazek_boost-invariant_1999,glazek_dynamics_2001,gomez-rocha_asymptotic_2015} reproduce the asymptotic-freedom result obtained from renormalization group techniques in Euclidean space~\cite{gross_ultraviolet_1973}.
%, at least for certain configurations of external particles. 
A finite dependence on the regularization functions used to regulate small momentum fractions (small-\textit{x}) usually remains~\cite{glazek_dynamics_2001,gomez-rocha_asymptotic_2015}. Such dependence needs further understanding.

A regularization provided by a canonical gluon mass~\cite{wilson_nonperturbative_1994} seems to be more adequate for various reasons\footnote{Regularization issues related to the introduction a gluon-mass parameter have been also considered in the context of other approaches to QCD (see e.g. Refs.~\cite{Cornwall:1979hz,Tissier:2011ey,Pelaez:2021tpq}).}: first of all, 
%ambiguity in the definition of the regularization functions for the canonical Hamiltonian is avoided because 
the same regulating function is used to remove both ultraviolet- and small-$x$ divergences;
furthermore, we use the same type of function as the ones introduced by the RGPEP procedure; and finally, it allows one to include a large range of +-component momenta near zero~\cite{glazek_massive_2019,glazek_computation_2020}. 
%finally, mass terms can be introduced consistently by using a similar procedure to the Higgs mechanism already tested for Abelian theories~\cite{glazek_massive_2019,glazek_computation_2020}, although not for non-Abelian ones yet because of the many complications inherent to these theories. 
In the following, we study the impact of introducing such a parameter and its consequences as a ragulator. At the end of the procedure, the limit of zero mass is applied, with no need of introducing new fields or interactions to recover gauge invariance. The result is qualitatively the same as the one obtained earlier~\cite{gomez-rocha_asymptotic_2015}: a function of the momentum fraction of external particles $h\left(x_{0}\right)$ appears as a side product of regularization and dumps asymptotic freedom for values of $x_{0}\lesssim0.13$.

This article is organized in the following way. In Section~\ref{FF} we present the basic elements involved in front-form quantization and the notation employed along this document. Section~\ref{RGPEP} is dedicated to introduce the RGPEP method and its application to the QCD Hamiltonian for gluons up to third order. It includes the regularization procedure. Section~\ref{coupling} defines the running coupling as a coefficient in the three-gluon-vertex Hamiltoian term. Finally, Section~\ref{results} concludes the article. 

%%%%%%%%%%%%%%%%%%%%%%%%%%%%%%%%%%%%%%%5%
\section{Front-Form Hamiltonian dynamics}
\label{FF}
%%%%%%%%%%%%%%%%%%%%%%%%%%%%%%%%%%%%%%%%%
Relativistic dynamics obeys the Poincaré algebra, a set of commutation relations between the ten fundamental dynamical quantities: the generators of space-time translations and rotations. In its original work~\cite{dirac_forms_1949}, Dirac found three ways of satisfying these relations, giving rise to the Instant-, Front-  and Point Forms of dynamics.

The RGPEP is built on the Front Form of dynamics for reasons we shall not discuss here (see the first sections of \cite{wilson_nonperturbative_1994}). In this form, four-vectors in Minkowski space are defined as $x^{\mu}=\left(x^{+},x^{-},x^{\perp}\right)$, where $x^+=x^0+x^3$, $x^-=x^0-x^3$, and $x^{\perp}=\left(x^{1},x^{2}\right)$. The inner product is \begin{equation}
    p\cdot q=p^{\mu}q^{\nu}g_{\mu\nu}=\frac{1}{2}p^{+}q^{-}+\frac{1}{2}p^{-}q^{+}-p^{\perp}\cdot q^{\perp} \ ,
\end{equation}
and
\begin{equation}
    p^{-} = \frac{p^{\perp2}+m^2}{p^{+}} \ , 
\end{equation}
represents the energy of the particle. 
The dynamics is not entirely specified by Dirac forms, and the Hamiltonian of interest is usually obtained from the $T^{+-}$ component of the energy-momentum tensor associated to the Lagrangian density considered. To describe pure-gluonic QCD we use the Yang-Mills theory of the non-Abelian gauge group SU(3). Details can be found in \cite{brodsky_quantum_1998,gomez-rocha_asymptotic_2015}, here we just quote the final expressions Eq. (9)-(14) of \cite{gomez-rocha_asymptotic_2015}:
\begin{equation}
    P^{-} = \frac{1}{2} \int_{\Omega} dx^{-}\,d^2 x^{\perp}\, \cH \ ,
\end{equation}
where $\cH$ is the Hamiltonian density and $\Omega$ denotes the surface of quantization, in this case, the plane defined by $x^{+}=\text{const}$. The Hamiltonian of pure-gluonic QCD has four terms
\begin{equation}
    \cH = \cH_{A^2}+\cH_{A^3}+\cH_{A^4}+\cH_{\left[\partial AA\right]^2} \ ,
    \label{CanonH}
\end{equation}
the subscripts on each of the four terms denote the number of fields involved in the term: $\cH_{A^2}$ is the free Hamiltonian, $\cH_{A^3}$ is the first-order vertex, $\cH_{A^4}$ is a four-gluon vertex and  $\cH_{\left[\partial AA\right]^2}$ appears due to the constraint equation
\begin{equation}
    A^{-}=2\frac{1}{\partial^{+}}\partial^{\perp}A^{\perp}-g\frac{2i}{\partial^{+2}}\left[\partial^{+}A^{\perp},A^{\perp}\right] \ ,
\end{equation}
in the gauge $A^{+}=0$. This sets $A^{-}=2\frac{1}{\partial^{+}}\partial^{\perp}A^{\perp}$ for free fields. The theory is quantized using the canonical expansion of field $A^{\mu}$ in terms of creation and annihilation operators with commutation relations
\begin{equation}
    \left[a_{k\sigma c},a^{\dagger}_{k'\sigma' c'}\right] = k^{+}\tilde{\delta}\left(k-k'\right)\delta^{\sigma \sigma'}\delta^{c c'} \ ,
\end{equation}
where $\sigma$ and \textit{c} are spin and color indices, respectively, and $\tilde{\delta}\left(p\right)=16\pi^3\delta\left(p^{+}\right)\delta\left(p^{1}\right)\delta\left(p^{2}\right)$. These relations and normal ordering of operators (denoted by $:H:$) are used to obtain the Hamiltonian in terms of creation and annihilation operators:
\begin{equation}
    H_{11} = \frac{1}{2}\int_{\Omega}d x^{-} d^2 x^{\perp}:\cH_{A^2}: \,\,=  \sum_{1}\int\left[1\right]\frac{k_{1}^{\perp 2}+\xi^2}{k_{1}^{+}}a^{\dagger}_{1}a_{1},
    \label{energyH}
\end{equation}
and 
\begin{align}
\begin{split}
    H_{21}+ &\,H_{12} = \frac{1}{2}\int_{\Omega}d x^{-} d^2 x^{\perp}:\cH_{A^3}:\\ 
    = & \,g\sum_{123}\int\left[123\right]f_{t_{r}}\tilde{\delta}\left(k^\dagger-k\right)Y_{123}a^{\dagger}_{1}a^{\dagger}_{2}a_{3}+h.c.,
\end{split}
    \label{firstvertex}
\end{align}
$\cH_{A^4}$ and $\cH_{\left[\partial AA\right]^2}$ give rise to Hamiltonian terms with four operators. The subscripts on the Hamiltonians denote the amount of creation and annihilation operators in the term, respectively; numbers 1, 2, 3 in sums and integrals refer to the respective degrees of freedom of particles 1, 2 and 3, e.g.,  $\left[123\right]=\left[k_{1}\right]\left[k_{2}\right]\left[k_{3}\right]$, and $k_i=dk^+_i dk^\perp_i/(16\pi^3k_i^+)$; the argument of the delta function $\left(k^{\dagger}-k\right)$ is a shortcut for the difference between momenta of created particles minus momenta of annihilated particles in the term. Finally, $Y_{123}$ is a polarization function whose concrete expression can be found in Eq.~(B3) of \cite{gomez-rocha_asymptotic_2015}. The parameter $\xi$ is the canonical gluon mass and $f_{t_{r}}$ is a regularization function, introduced in the next section.  The subscript  $t_r$ is a cutoff parameter. 
\begin{figure*}
\centering
\includegraphics[scale=0.40]{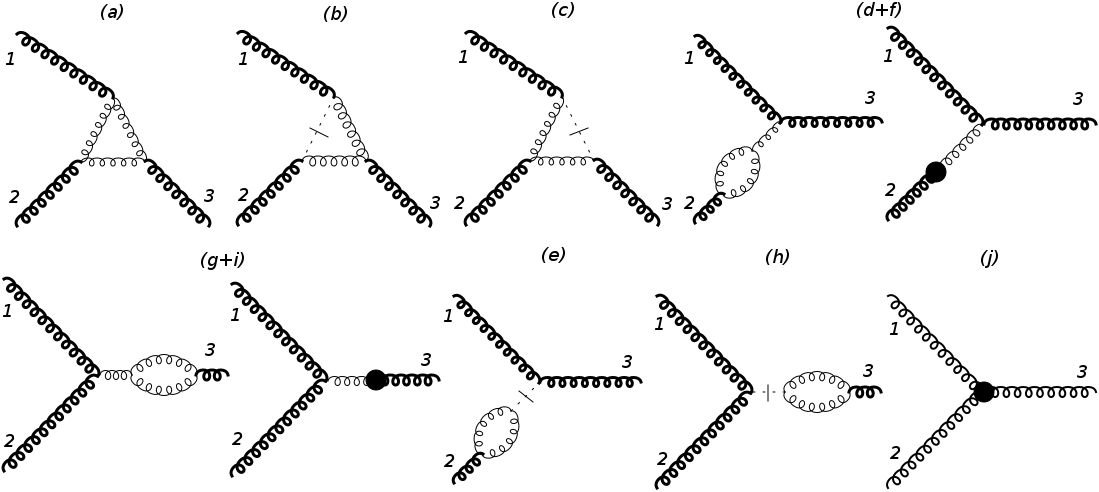}
\caption{Third-order contributions to the running coupling, including the counterterm. Terms (\textit{a})-(\textit{i}) correspond to $\gamma\left(a\right)-\gamma\left(i\right)$ of Eq.~(\ref{definitiongammas}), term (\textit{j}) is the third-order counterterm corresponding to $\gamma\left(j\right)$. External \textit{effective} particles are labeled 1, 2 and 3 and are represented with bold gluonic lines.}
\label{complete3gvertex}       % Give a unique label
\end{figure*}

%%%%%%%%%%%%%%%%
\section{Renormalization Group Procedure for Effective Particles}
\label{RGPEP}
%%%%%%%%%%%%%%%%5
Canonical expressions with regulators such as Eq.~(\ref{firstvertex}) are transformed in order to produce results independent of regularization. RGPEP takes them as initial conditions and sets a family of equivalent Hamiltonians that depend on a parameter $t$:
\begin{equation}
    H_{0}\left(a_{0}\right) = H_{t}\left(a_{t}\right),
    \label{scalet}
\end{equation}
the new operators $a_{t}$ create and annihilate effective particles of size \textit{$s=\sqrt[4]{t}$} and are related to the initial or bare operators by a unitary transformation (cf. Ref. \cite{glazek_perturbative_2012})
\begin{equation}
    a_{t} = \cU_{t}a_{0}\cU_{t}^{\dagger},
\end{equation}
whose anti-hermitian generator is 
\begin{equation}
    \cG_{t} = \left[H_{f},H_{Pt}\right] \ .
\end{equation}
This expression,  $H_{f}$ is the free part of the Hamiltonian which does not evolve with \textit{t}; $H_{Pt}$ is the final Hamiltonian multiplied by half the sum of total momentum created and annihilated in that term. The generator gives rise to a  differential equation with a double-commutator of the type of that introduced by Wegner~\cite{wegner_flow-equations_1994}:
\begin{equation}
    \frac{d H_{t}}{dt} = \left[\left[H_{f},H_{Pt}\right],H_{t}\right] \ .
    \label{MasterRGPEP}
\end{equation}
Note that Eq.~(\ref{scalet}) forces functions multiplying operators in Hamiltonian terms to also change with \textit{t}. In order to distinguish the change of these functions with the change of operators we will use normal fonts $H_{t}\left(a_{t}\right)$ when both are at the scale \textit{t} and calligraphic font $\cH_{t}\left(a_{0}\right)$ when the operators are at the bare scale. 

Eq.~(\ref{MasterRGPEP}) can be solved order by order in a perturbative expansion on the coupling constant \textit{g}. Taking into account only those terms relevant to the derivation of the running coupling, Eq.~(27)-(33) of \cite{gomez-rocha_asymptotic_2015} we have:
\begin{align}
\begin{split}
    H_{t} = & \,H_{11,0,t} + H_{21,g,t}+H_{12,g,t}\\
    &+H_{11,g^2,t}+H_{22,g^2,t}+H_{31,g^2,t}+H_{13,g^2,t}\\
    &+H_{21,g^3,t}+H_{12,g^3,t},
\end{split}
\end{align}
in order to alleviate notation and build intuition we follow \cite{gomez-rocha_asymptotic_2015} and define
\begin{align}
    H_{11,0}&\rightarrow \cE,\\
    H_{11,g^2}&\rightarrow g^2 \hat{\mu}^2,\\
    H_{22,g^2}&\rightarrow g^2 \cX_{22},\\
    H_{31,g^2}+H_{13,g^2}&\rightarrow g^2\Xi_{31}+g^2\Xi_{13},\\
    H_{21,g}+H_{12,g}&\rightarrow g\cY_{21}+g \cY_{12},\\
    H_{12,g^3}+H_{21,g^3}&\rightarrow g^3 \cK_{21} + g^3 \cK_{12},
\end{align}
These expressions are then introduced in Eq.~(\ref{MasterRGPEP}) and give rise to successive expressions in powers of $g$. Counterterms are introduced order by order in the initial Hamiltonian to make physical results independent of regularization.

%%%%%
\subsection{First-order solution}
%%%%%
Let us introduce some important concepts before analyzing the three-gluon vertex and the running coupling. The equation in first power of $g$ has two terms: one corresponding to $\cY_{21t}$, the other to its Hermitian conjugate $\cY_{12t}$. For the first one we have
\begin{equation}
    \cY_{21t}' = \left[\left[\cE,\cY_{21Pt}\right],\cE\right].
\end{equation}
\begin{figure}
    \centering
    \includegraphics[scale=0.7]{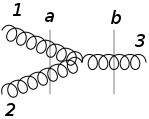}
    \caption{Diagrammatic representation of $\cY_{21t}$. Letters denote configurations of particles before and after the interaction, \textit{a} refers to particles 1 and 2 and \textit{b} to particle 3. For more details about notation see Ref.~\cite{glazek_perturbative_2012}. }
    \label{three-gluon-vertex}
\end{figure}
The solution $\cY_{21}$ is represented graphically by figure \ref{three-gluon-vertex} and it is similar to Eq.~(\ref{firstvertex}) 
\begin{equation}
    \cY_{21t} = g\sum_{123}\int\left[123\right]f_{t,ab}f_{t_{r},ab}\tilde{\delta}\left(k^\dagger-k\right)Y_{123}a^{\dagger}_{1}a^{\dagger}_{2} a_{3}   \ ,
\label{Y21t}
\end{equation}
with
\begin{equation}
    f_{t,ab} = \exp\left[-t\left(\cM_{a}^2-\cM_{b}^2\right)^2\right] \ ,
    \label{formfactor}
\end{equation}
where $\cM_{i}$ is the invariant mass of configuration \textit{i}. In this case we have: 
\begin{align}
    \cM_{a}^2 =&  \,\frac{\kappa_{12}^{\perp 2}+\xi^2}{x_{1/3}x_{2/3}} \ ,\\
    \cM_{b}^2 = &\,\xi^2 \  ,
\end{align}
$x_{1/3}$ and $x_{2/3}$ are the longitudinal momentum fractions of particles 1 and 2, respectively, and $\kappa_{12}^{\perp}$ is the relative transverse momentum of particles in configuration \textit{a}. More generally,  we call \textit{parent momenta P} to the sum of momenta created or annihilated through a given interaction, the longitudinal momentum fraction of particle \textit{p} involved in such interaction is then 
\begin{equation}
    x_{p/P} = p^{+}/P^{+}, 
\end{equation}
and the transverse momentum is 
\begin{equation}
    \kappa_{p/P} = p^{\perp}-x_{p/P}P^{\perp} \ ,
\end{equation}
corresponding to figure~\ref{three-gluon-vertex} we have $P=p_{3}$ and $\kappa_{12}^{\perp} = \kappa_{1/3}^{\perp}=-\kappa_{2/3}^{\perp}$.

Eq.~(\ref{Y21t}) justifies the name \textit{effective particles of size} $s = \sqrt[4]{t}$. Namely, form factors like Eq.~(\ref{formfactor}) prevent
particles of size $s$ to change their relative kinetic energy by more than about $\lambda=1/s$ through a single interaction. 
Note that the notion of \textit{size} is inherent to interactions; momenta of free particles are not constrained in this formalism no matter the value of \textit{s}.

Finally, the canonical expressions are regularized by the introduction of a canonical gluon mass $\xi$ and a regulating function defined through Eq.~(\ref{formfactor}) $t=t_{r}$, where $t_r$ is a small value that acts as a cutoff. 
Frequently, the notation $f_{t+t_{r},ab}$ is used instead of $f_{t,ab}f_{t_{r},ab}$, since it allows to clearly see that for any finite value of \textit{t} the regularization parameter is ``muted'' in the limit $t_{r}\rightarrow0$ \cite{glazek_computation_2020}.

%%%%%%%%%%%
\subsection{The three-gluon vertex}
\label{3g-vertex}
%%%%%%%%%%%%
The three-gluon vertex can be analyzed by considering the third-order solution to the RGPEP equation and it has the following structure
\begin{align}
\begin{split}
\cV_{21t}=g\cY_{21t}+g^3\cK_{21t}
&=\left(\sum\int\right)_{123}\tilde{\delta}\left(p^{\dagger}-p\right)f_{t+t_{r},ab}\\
&\times\left\lbrace g\tilde{\cY}_{21t}+g^3\left(\tilde{\cK}_{21t}+\tilde{\cK}_{210}\right)\right\rbrace
\end{split}
\label{3orgvertex}
\end{align}
where $\cY_{21,t}$ and $\cK_{21,t}$ are the first and third order contributions respectively. Caligraphic letters with tildes are introduced to make explicit the common factors within integrals. $\tilde{\cK}_{21,0}$ is the third order counterterm. We focus on terms which can be factorized in the following way:
\begin{equation}
    \tilde{\mathcal{K}}_{21t}\left(x_{1},\kappa_{12},\sigma\right)=c_{t}\left(x_{1},\kappa_{12}\right)Y_{123}\left(x_{1},\kappa_{12},\sigma\right),
\end{equation}
where $Y_{123}\left(x_{1},\kappa_{12},\sigma\right)$ is the canonical spin and color structure of the first-order interaction of the initial Hamiltonian. $c_{t}$ is the function obtained from the RGPEP procedure that multiplies the operator structure defining the three gluon vertex, i.e. $a^\dagger a^\dagger a+h.c.$ It can be written as the sum of diagrams \textit{a} to \textit{i} of figure~\ref{complete3gvertex}, denoted by $\gamma\left(a\right),\gamma\left(b\right),...,\gamma\left(i\right)$:
\begin{equation}
\tilde{\mathcal{K}}_{21t}=c_{t}\left(x_{0},\kappa_{12}\right)=\frac{\sum_{n}\gamma\left(n\right)}{2\cdot16\pi^{3}} \ ,
\label{definitiongammas}
\end{equation}
each one of these functions involve three-dimensional loop integrals characterized by the Front-Form momentum fractions \textit{x} and relative transverse momenta $\kappa^\perp$ of the internal virtual particles, and would diverge in the limits $\kappa\rightarrow\infty$ and $x\rightarrow0,1$ in the absence of form factors and regulators. 

%%%%%%%%%%%
\subsection{Regularization}
%%%%%%%%%%%
\label{Reg}
RGPEP form factors suppress interactions if the differences of invariant masses $\left(\cM^2_f - \cM^2_i\ , \ \cM^2=\frac{\kappa^2+m^2}{x(1-x)}\right)$ between the initial and final states in a given interaction are greater than the effective size parameter $s=\sqrt[4]{t}$, and thus indirectly avoid the appearance of large $\kappa$ divergences. However, the regularization is incomplete: at $t=0$ the effective expressions must reduce to the ones of the initial theory, which translates to differences such as $f_{t}-1$, with $f_{t}$ a form factor and $f_{0} = 1$. Contributions coming from $-1$ factors are not regularized and give rise to loop divergences. To avoid such divergences we introduce functions $f_{t_{r}}$ in the initial Hamiltonian $\cH_{0}$.
%, replacing the $-1$ for $-f_{t_{r}}$ in the RGPEP solutions.

Counterterms are necessary to avoid dependence on the regularization factor $t_{r}$ in physical results. To find them we notice that the effective Hamiltonians $\cH_{t}$ become independent of $t$ in the ultraviolet limit $\kappa\rightarrow\infty$, and only form factors with vanishingly small $t_{r}$ remain. Thus the difference between two scales $\cH_{t}-\cH_{t_{0}}$ is ultraviolet finite regardless the values of $t$ and $t_{0}$. The ultraviolet divergent part of the counterterm can then be considered to be that of $-\cH_{t_{0}}$, and its finite $\left(\text{in the limit } \kappa\rightarrow\infty\right)$ part should be then fixed by experimental considerations, for more details see \cite{glazek_boost-invariant_1999,glazek_dynamics_2001}.

We have now justified the following equation for the third order counterterm:
\begin{equation}
\tilde{\cK}_{210} = -\left(c_{t_0}\left(x_{1},\kappa_{12}\right)-c_{0}\left(x_1,\kappa_{12},\sigma\right)\right)Y_{123}\left(x_1,\kappa_{12},\sigma\right),
\label{3counterterm}
\end{equation}
where the function $c_{t_{0}}$ is the same that we introduced in Eq.~$\left(\ref{definitiongammas}\right)$ with $t$ changed to $t_{0}$, $c_{0}$ is a finite and in principle unknown contribution necessary because in general the finite part of the counterterm is not equal to the finite part of $c_{t_{0}}$.

The situation for small-\textit{x} divergences ($x\rightarrow0,1$) is somehow different: In the massless case, invariant masses remain finite if $\kappa\rightarrow0$ in addition, avoiding form factors to regulate these $x$ divergences. Several strategies are now possible: in \cite{glazek_dynamics_2001,gomez-rocha_asymptotic_2015} one introduces different regularization functions and considers the impact of their choice in the running coupling. Here, in contrast, a gluon mass $\xi$ and initial functions $f_{t_{r}}$ are used. With a gluon mass invariant masses diverge if momentum fractions $x$ approach their limiting values for any $\kappa$, and thus form factors avoid also these divergences. It is still necessary to consider a parameter $t_{r}$ different from zero, but we do not need extra regularization functions whose explicit forms are in principle arbitrary. At the end of the procedure we take the limit $\xi\rightarrow0$ to recover QCD massless gluons.

%%%%%%%%%%%%%%%%%%%%%%%%%%
\section{Running coupling}
\label{coupling}
%%%%%%%%%%%%%%%%%%%%%%%%%%
We use the definition of the running coupling introduced in \cite{glazek_dynamics_2001,gomez-rocha_asymptotic_2015}: \textit{the running coupling is  defined as the coefficient in front of the canonical color, spin and momentum dependent factor $Y_{123}\left(x_1,\kappa_{12},\sigma\right)$ in the limit $\kappa_{12}\rightarrow0$ for some value of $x_{1}$ denoted $x_{0}$}.
Therefore, we first factorize the function $Y_{123}\left(x_1,\kappa_{12},\sigma\right)$ in Eq.~(\ref{3orgvertex}):
\begin{equation}
\begin{split}
&\tilde{\mathcal{V}}_{21t}\left(x_1,\kappa_{12},\sigma\right)=Y_{123}\left(x_1,\kappa_{12},\sigma\right)\\ &\times\left\{g+g^{3}\left[c_{t}\left(x_1,\kappa_{12}\right)-c_{t_{0}}\left(x_1,\kappa_{12}\right)-c_{0}\left(x_1,\kappa_{12},\sigma\right)\right]\right\}\ .
\end{split}
\label{factorizationY}
\end{equation}
By definition, the running coupling reads
\begin{equation}
g_t=g+g^3\lim_{\kappa_{12}\rightarrow 0}\left[c_{t}\left(x_{0},\kappa_{12}\right)-c_{t_{0}}\left(x_{0},\kappa_{12}\right)-c_{0}\left(x_{0},\kappa_{12},\sigma\right)\right],
\end{equation}
setting its value to be $g_0$ at the scale $t_0$, one has
\begin{equation}
g_t=g_0+g^3_0\lim_{\kappa_{12}\rightarrow0}\left[c_{t}\left(x_{0},\kappa_{12}\right)-c_{t_{0}}\left(x_{0},\kappa_{12}\right)\right] \ ,
\label{running-coupling}
\end{equation}
where
\begin{equation}
c_{t}\left(x_{0},\kappa_{12}\right)=\frac{\sum_{n}\gamma\left(n\right)}{2\cdot16\pi^{3}} \  ,
\label{ct}
\end{equation}
and \textit{n} runs from \textit{a} to \textit{i}. Eq.~(\ref{running-coupling}) can now be written in terms of the difference of $\gamma$s at scales $t$ and $t_{0}$:
\begin{equation}
    g_{t}=g_{0}+g_{0}^3\lim_{\kappa_{12}\rightarrow0}\frac{\sum_{n}\left(\gamma_{t}\left(n\right)-\gamma_{t_{0}}\left(n\right)\right)}{2\cdot16\pi^3} \ .
\end{equation}

Explicit expressions for $\gamma$s can be obtained from  Appendix~C of \cite{gomez-rocha_asymptotic_2015}, changing the RGPEP factors $\cB_{t}$ as described in appendix \ref{massivegluons}. These equations usually involve integrals in momentum fraction $x$ and relative transverse momenta $\kappa$ of internal virtual particles. They are evaluated as explained in Appendix~\ref{integrationappendix}. Finally, relevant results are obtained after applying limits $\xi\rightarrow0$ and $t_{r}\rightarrow0$. 

\subsection{Term \textit{a}}
The triangle term $a$ is obtained from the product of three first-order vertices $\cY_{t}$. Introducing (barred) dimensionless variables defined in  Eq.~(\ref{dimensionless}), we can express it as
\begin{align}
\begin{split}
    \gamma_{t}\left(a\right)-\gamma_{t_{0}}\left(a\right)
    \ = \ 
    &N_{c}\pi\log\left(\frac{t}{t_{0}}\right)\left[-\frac{11}{3}+\frac{1}{6}h_{a}\left(x_{1}\right)\right]\\
    &-\frac{16\pi N_{c}}{x_1x_2 } \frac{\bar{t}-\bar{t}_{0}}{x_{1}^2+x_{2}^2}
    \frac{\bar{\xi}}{\bar{t}_{r}} \  ,
\end{split}
\label{terma}
\end{align}
where 
\begin{align}
\begin{split}
&\frac{1}{6}h_{a}\left(x_{1}\right)=-3\log\left(\bar{\xi}^{4}\sqrt{\bar{t}\bar{t}_{1}}e^{\gamma_{E}}\right)-5-\frac{2}{1-x_{2}^{2}}\log\left(\frac{1+x_{2}^{2}}{x_{1}x_{2}}\right)\\
&-\frac{2}{1-x_{1}^{2}}\log\left(\frac{1+x_{1}^{2}}{x_{1}x_{2}}\right)-\log\left(2\right)+\frac{1-x_{1}^2x_{2}^2}{\left(1+x_{1}^2\right)\left(1+x_{2}^2\right)}\\
&+\left(1-\frac{1}{1-x_{1}^{2}}-\frac{1}{1-x_{2}^{2}}\right)\log\left[\frac{\left(x_{1}^{2}+x_{2}^{2}\right)x_{1}^{2}x_{2}^{2}}{2\left(1+x_{2}^{2}\right)\left(1+x_{1}^{2}\right)}\right],
\end{split}
\label{ha}
\end{align}
with $x_{2}=1-x_{1}$, and $\gamma_{E}$ the Euler-Mascheroni constant.
%$\xi$ is the gluon mass, 

\subsection{Term \textit{b}}
Term $b$ is obtained from the product of the first-order vertex $\cY_{t}$ and the second-order term $\cX_{t}$
\begin{equation}
\gamma_{t}\left(b\right)-\gamma_{t_{0}}\left(b\right)
\ = \  
\frac{16\pi N_{c}}{x_{0}x_{2}}\frac{\left(\bar{t}-\bar{t}_{0}\right)}{x_{1}^{2}+x_{2}^{2}}\frac{\bar{\xi}}{\bar{t}_{r}}.
\label{termb}
\end{equation}
this contribution exactly cancels the term proportional to $\bar{t}-\bar{t}_{0}$ in Eq.~(\ref{terma}).

\subsection{Terms \textit{d} and \textit{f}}
Term $d$ is obtained from the product of the second order self-energy term $\hat{\mu}_{t}$ and the first-order vertex $\cY_{t}$; while term $f$ from the second-order counterterm and the first order vertex $\cY_{t}$. Their sum gives the following result
\begin{equation}
    \gamma_{t}\left(d\right)-\gamma_{t_{0}}\left(d\right) \ = \ \pi N_{c}\log\left(\frac{t}{t_{0}}\right)\left(\frac{11}{3}+\frac{1}{6}h_{d+f}\left(x_{1}\right)\right) \ ,
\label{termdf}
\end{equation}
where 
\begin{align}
    &\frac{1}{6}h_{d+f}\left(x_{1}\right) \ = \ 2\log\left(e^{\gamma_{E}}\bar{\xi}^{4}\sqrt{\bar{t}\bar{t}_{0}}\right)+2\log{2}+4 \nonumber \\
    &- 2\frac{x_{2}^{2}}{1-x_{2}^{2}}\log\left(\frac{1+x_{2}^{2}}{2x_{2}^{2}}\right)-2\frac{x_{1}^{2}}{1-x_{1}^{2}}\log\left(\frac{1+x_{1}^{2}}{2x_{1}^{2}}\right) \  .
\label{hdf}
\end{align}

\subsection{Terms \textit{g} and \textit{i}}
Terms $g$ and $i$ are obtained in a similar way that terms $d$ and $f$:
\begin{equation}
\gamma_{t}\left(g+i\right)-\gamma_{t_{0}}\left(g+i\right)
\ = \ 
\frac{\pi N_{c}}{6}\log\left(\frac{t}{t_{0}}\right)\left(11+h_{g+i}\left(x_{1}\right)\right) \ ,
\label{termgi}
\end{equation}
with
\begin{equation}
\frac{1}{6}h_{g+i}\left(x_{1}\right) \ = \ \log\left(e^{\gamma_{E}}\bar{\xi}^4\sqrt{\bar{t}\bar{t_{0}}}\right)+\log{2}+1\ .
\label{hgi}
\end{equation}

\subsection{Terms \textit{c}, \textit{e} and \textit{h}}
Term $c$ is obtained from the product of the first-order vertex $\cY_{t}$ and the second-order interaction $\Xi_{t}$. The result turns out to be negligible in the limits $\xi\rightarrow0$ and $t_{r}\rightarrow0$. Terms \textit{e} and \textit{h} are also derived from the same vertices, and do not contribute to the running coupling since there are no linear terms in $\kappa_{12}$ that could give rise to the canonical polarization structure $Y_{123}$ of Eq.~(\ref{factorizationY}).

%%%%%%%%%%%%%%%%%%%%%%%%%%%%%%%%%
\section{Results and conclusions}
\label{results}
\begin{figure}
\centering
\includegraphics[width=8cm,clip]{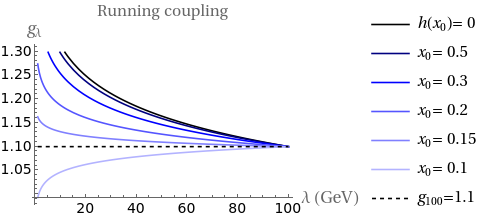}
\caption{Running coupling for different values of $x_{0}$, the black line $(h\left(x_{0}\right)=0)$ represents the result obtained from the renormalization group equations in Euclidean space. The function exhibits asymptotic freedom from $x_0 = 0.5$ down to values of $x_{0}\approx0.13$.}
\label{Rc_plot_total}      % Give a unique label
\end{figure}

%%%%%%%%%%%%%%%%%%%%%%%%%%%%%%%%%
Eqs.~(\ref{ha}), (\ref{hdf}) and~(\ref{hgi}) give the final expression for the running coupling constant
\begin{equation}
    g_{t} = g_{0}-N_{c}\frac{g_{0}^{3}}{48\pi^{2}}\log\left(\frac{\lambda}{\lambda_{0}}\right)\left[11+h\left(x_{1}\right)\right],
\label{Rcfinal}
\end{equation}
with $\lambda = 1/\sqrt[4]{t}$ and
\begin{align}
\begin{split}
&h\left(x_{1}\right) =-6\left\{\frac{1+x_{2}^{2}}{1-x_{2}^{2}}\log\left[\frac{\left(1+x_{2}^{2}\right)^{2}}{x_{2}^{2}}\right]\right.\\
&+\frac{1+x_{1}^{2}}{1-x_{1}^{2}}\log\left[\frac{\left(1+x_{1}^{2}\right)^{2}}{x_{1}^{2}}\right]-\frac{1-x_{1}^2x_{2}^2}{\left(1+x_{1}^2\right)\left(1+x_{2}^2\right)}\\
&\left.+\left(1-\frac{1}{1-x_{1}^{2}}-\frac{1}{1-x_{2}^{2}}\right)\log\left[\frac{8\left(1+x_{2}^{2}\right)\left(1+x_{1}^{2}\right)}{\left(x_{1}^{2}+x_{2}^{2}\right)}\right]\right\}.
\end{split}
\end{align}
Eq. (\ref{Rcfinal}) is represented in figure~\ref{Rc_plot_total} for values of $x_1=x_{0}$ ranging from 0.5 to 0.1. The result exhibits asymptotic freedom for $x_{0}$ down to 0.13 and it coincides with the analysis of Feynman diagrams in pure gluonic QCD \cite{gross_ultraviolet_1973} if the factor $\lambda$ is interpreted as the scale of the renormalization group equations in Euclidean space and if $h\left(x_{0}\right)=0$  (cf.~\cite{glazek_dynamics_2001,gomez-rocha_asymptotic_2015}).

\begin{figure*}
\includegraphics[scale=0.55]{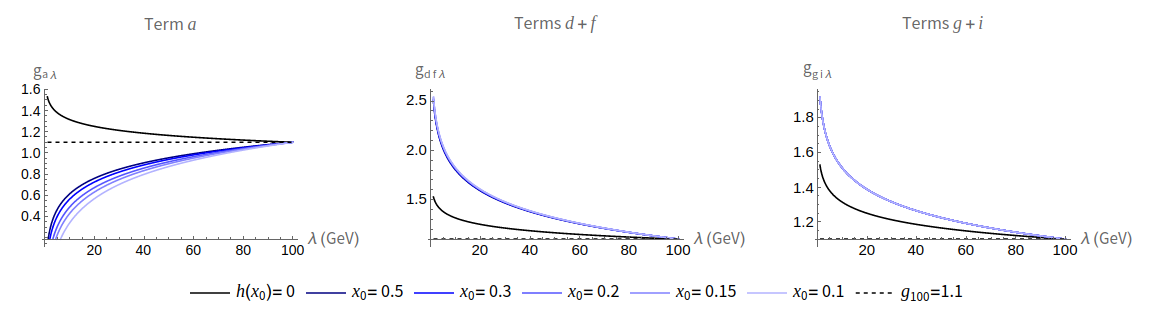}
\caption{Relevant contributions from terms $a$, $d+f$ and $g+i$ to the running coupling (terms linear in the difference $\bar{t}-\bar{t}_{0}$ and logarithms $\log\left(e^{\gamma_{E}}\sqrt{\bar{t}\bar{t}_{0}}\bar{\xi}^4\right)$ are not taken into account because they cancel in the final expression). Self-energy terms contribute to asymptotic freedom,  the triangle term does not, and dominates over the other two for low values of $x_{0}$.}
\label{Rc_separate}
\end{figure*}

In figure~\ref{Rc_separate} the contributions from different terms are considered separately. The self-energy ones, corresponding to \textit{d+f} and \textit{g+i} increase as the energy scale diminishes and thus contribute to asymptotic freedom. In contrast, \textit{a} decreases with the energy scale, and thus the loss of asymptotic freedom for low values of $x_{0}$ is entirely due to the triangle term \textit{a}.

There is no dependence on the mass parameter $\xi$ in the final result in the limit $\xi\rightarrow0$ even though separate contributions diverge in this limit. Thus a mass term for gluons seem to provide an adequate regularization of small-$x$ divergences, producing a function of $h\left(x_{1}\right)$ that controls the strength of the running of the coupling constant for different values of the external longitudinal momentum fraction, with the same qualitative behaviour obtained in \cite{glazek_dynamics_2001,gomez-rocha_asymptotic_2015}. Finally, as noted in appendix \ref{integrationappendix}, the methods developed here can also be used to evaluate fermion integrals when particles' masses are small compared to the scales settled by $t$ and $t_{0}$.

\section*{Acknowledgements}
We thank Professor Stanisław D. Głazek for fruitfull discussions and acknowledge financial support from the FEDER funds, project ref. A-FQM-406-UGR20 and from MCIN/ AEI/10.13039/501100011033, Project Ref. PID2020-114 767GB-I00.
Figures~\ref{complete3gvertex} and~\ref{three-gluon-vertex} are made using the open software JaxoDraw~\cite{binosi_jaxodraw_2004} distributed under the GNU General Public license.
\appendix
\section{Introduction of a mass term for gluons}
\label{massivegluons}
As described in Subsection~\ref{3g-vertex} each $\gamma\left(n\right)$ consists on a three dimensional integral over momentum fractions $x$ and relative transverse momenta $\kappa$ of internal virtual particles. RGPEP factors in integrands depend on the order in the perturbation expansion, on how these particles are connected, and on polarization functions that encode the spin and color of these internal degrees of freedom. In the case of massless gluons, explicit expressions for these factors are found in Appendix C of~\cite{gomez-rocha_asymptotic_2015}. The addition of a gluon mass alter these equations, changing invariant masses that appear there for:
\begin{equation}
    \cM_{\chi,ij}^2 =  \cM_{ij}^2-\xi^2 = \frac{\kappa_{ij}^2+\xi^2\left(1-x_i x_j\right)}{x_ix_j}:=\frac{\kappa_{ij}^2+\chi_{ij}^2}{x_i x_j},
\end{equation}
\begin{equation}
    \cM_{\chi,16}^2 =  \cM_{16}^2-\xi^2 =  x_{1}\frac{\kappa^{2}+\xi^{2}\left(\frac{x^{2}-x_{1}x_{6}}{x_{1}^{2}}\right)}{\left(x-x_{1}\right)} := x_{1}\frac{\kappa^{2}+\chi_{16}^{2}}{\left(x-x_{1}\right)},
\end{equation}
\begin{equation}
    \cM_{\chi,168}-\xi^2=\frac{\cM_{\chi,68}^2}{x_2}+\cM_{\chi,12}^2,
\end{equation}
with ij= $\left\{68, 78, 12\right\}$. 
Note that they may be regarded as invariant masses $\cM_{\chi}$ with $x$-dependent ``masses'' $\chi\left(x\right)$.
Numbers denote variables of particles in the various interactions of the third-order diagrams of figure~\ref{complete3gvertex}. 
\section{Integration method}
\label{integrationappendix}
To evaluate the expressions that are solutions to the RGPEP equation we use dimensionless variables:
\begin{equation}
\bar{t}=\frac{t}{t_{N}},\,\bar{\kappa}^{\perp}=\kappa^{\perp}t_{N}^{1/4},\,\bar{\xi}=\xi t_{N}^{1/4} \ , 
\label{dimensionless}
\end{equation}
where $t_{N}$ is an arbitrary scale. The integrals over momentum fractions \textit{x} are then divided in three intervals or regions
\begin{equation}
    \lim_{\bar{\xi}\rightarrow 0} \left[\int_{0}^{\bar{\xi}}+\int_{\bar{\xi}}^{1-\bar{\xi}}+\int_{1-\bar{\xi}}^{1}\right] dx \int_{\bar{\xi}}^{\infty}  d^2\bar{\kappa}^{\perp} G\left(x,\bar{\kappa}^{\perp},\bar{t},\bar{t}_{0};\bar{t}_{r},\bar{\xi}\right),
\end{equation}
called region $\mathbf{I}$, region $\mathbf{II}$, and region $\mathbf{III}$ respectively; $G\left(x,\bar{\kappa}^{\perp},\bar{t},\bar{t}_{0};\bar{t}_{r},\bar{\xi}\right)$ is usually a function of invariant masses, form factors and polarization vectors that is simplified as follows: 
\begin{itemize}
    \item In region $\mathbf{II}$ the polarization fraction \textit{x} is bounded $\bar{\xi}<x<1-\bar{\xi}$ and no integral diverges because the ultraviolet $\kappa\rightarrow\infty$ have been already regularized. Thus we set the regularization parameters to zero in the integrand: $G\left(x,\bar{\kappa}^{\perp},\bar{t},\bar{t}_{0};0,0\right)=G\left(x,\bar{\kappa}^{\perp},\bar{t},\bar{t}_{0};\bar{t}_{r},\bar{\xi}\right)|_{\mathbf{II}}$ and apply Eq.~(E17) of \cite{gomez-rocha_asymptotic_2015}
\begin{equation}
    \int d^2\bar{\kappa}^\perp\frac{f_{t}-f_{t_{0}}}{\bar{\kappa}^{\perp2}} = \frac{\pi}{2}\ln\frac{\bar{t_{0}}}{\bar{t}}.
\end{equation}
 Integrals over \textit{x} are then easily evaluated and only divergent and constant terms in the limit $\xi\rightarrow0$ are kept.

\item 
Region $\mathbf{I}$ is more involved because invariant masses do diverge even in the limit $\xi\rightarrow0$. Nevertheless, since $\textit{x}<\bar{\xi}$, it is enough to factorize the poles in $x=0$ and expand around this point the remaining terms to obtain the most strongly-divergent results. For example, a typical integral to evaluate would be
\begin{equation}
    \int_{0}^{1}dx\frac{1}{x\left(1-x\right)}\left[\Gamma\left(0,\frac{\alpha\left(x\right)\, \bar{t}\,\bar{\xi}^4}{x^2\left(1-x\right)
    ^2}\right)-\Gamma\left(0,\frac{\alpha\left(x\right)\, \bar{t}_{0}\,\bar{\xi}^4}{x^2\left(1-x\right)
    ^2}\right)\right],
\end{equation}
where $\Gamma\left(0,x\right)$ is the incomplete gamma function and $\alpha\left(x\right)$ is finite in $x=0$ and $x=1$. In region $\mathbf{I}$ we can evaluate the main contribution in the limit $\xi\rightarrow 0$ by considering
\begin{equation}
\int_{0}^{\bar{\xi}}dx\frac{1}{x}\left[\Gamma\left(0,\frac{\alpha\left(0\right)\, \bar{t}\,\bar{\xi}^4}{x^2}\right)-\Gamma\left(0,\frac{\alpha\left(0\right)\, \bar{t}_{0}\,\bar{\xi}^4}{x^2}\right)\right],
\end{equation}
which yields
\begin{equation}
    \frac{1}{4}\log\left(\frac{\bar{t}}{\bar{t}_{0}}\right)\left(2\log\left(e^{\gamma_{E}}\right)+2\log\left(\alpha\right)+\log\left(\bar{t}\bar{\xi}^{4}\right)\right)+\mathcal{O}\left(\bar{\xi}^2\right).
\label{gamma-integral}
\end{equation}
The cutoff $t_{r}$ usually appears added to $t$ or $t_{0}$ in form factors, $f_{t+t_{r}}$ and $f_{t_{0}+t_{r}}$; in these cases it is ``muted'' and can be discarded. However, special care should be taken when this is not the case, as there are contributions depending on $t_{r}$ in equations Eq.~(\ref{terma}) and Eq.~(\ref{termb}).
Region $\mathbf{I}$ is different for the triangle terms $a$, $b$ and $c$, since the low limit of integration over $x$ changes from zero to $x_{1}$. In these cases the poles at $x_{1}$ are factorized instead and the remaining expressions expanded around this point.
\item
For terms $d-i$ of figure \ref{complete3gvertex} results of region $\mathbf{I}$ can also be applied to region $\mathbf{III}$ because the integrals are symmetric under the change of variables $y=1-x$.
For triangle terms $a-c$ the simplification  of region $\mathbf{I}$ can be applied factorizing poles in $x=1$ and expanding around this point instead of $x=x_{1}$.
\end{itemize}
Finally, contributions of light fermions beyond the ultraviolet counterterm already found in~\cite{glazek_dynamics_2001}  can be evaluated using this method replacing the gluons mass parameter $\xi$ with the fermion mass $m_{f}$ if the scales settled by the parameters $t$ and $t_{0}$ are much greater than $m_{f}$.
\bibliography{ProceedingsConfXV}
\end{document}